# Flow induced two-dimensional ζ potential on chemically uniform solid-liquid interface: A possible electrokinetic origin of streamwise vortices

**Shuangshuang Meng, [1, #] Yu Han, [1, #], Xiaoqiang Feng, [1] Yueqiang Zhu, [1] Jin'an Pang, [1] Duyang Zang, [2] Yu Chen, [1] Ruihong Zhang, [1] Chen Zhang, [1] Ce Zhang, [1, *] Kaige Wang, [1, *] Jintao Bai, [1] and Wei Zhao [1, *]**

[1] State Key Laboratory of Photon-Technology in Western China Energy, International Collaborative Center on Photoelectric Technology and Nano Functional Materials, Institute of Photonics & Photon Technology, Northwest University, Xi'an 710127, China

[2] School of Physical Science and Technology, Northwestern Polytechnical University, Xi'an 710129, P. R. China

* Corresponding author: zhangce@nwu.edu.cn; wangkg@nwu.edu.cn; zwbayern@nwu.edu.cn

# These authors contribute equally to this investigation

**Abstract** The ζ potential at solid-liquid interfaces governs electrokinetic phenomena across chemistry, biology, and engineering. Despite its centrality, local measurement of this potential has remained a formidable challenge, forcing the universal assumption of spatial uniformity on chemically homogeneous surfaces. Here we introduce Fluorescence Photobleaching Electrochemistry Analysis (FLEA)—a non-invasive technique that achieves 180 nm lateral resolution—to map the ζ potential at unprecedented spatial detail. We discover that a perfectly uniform glass surface, under laminar flow, develops an intrinsic two-dimensional ζ potential landscape: approximately symmetrical V-shaped distribution across the channel width that varies systematically with flow velocity, pH, and ionic strength. This two-dimensional electrochemical heterogeneity generates a rotational electric body force that spontaneously produces streamwise vortices—structures commonly observed in boundary layers. Our findings establish FLEA as a transformative tool for surface characterization, reveal a previously unrecognized origin of streamwise vorticity at interfaces from the aspect of electrokinetic.

## 1. Introduction

The electric double layer (EDL)—the nanoscopic ionic atmosphere that forms at every solid-liquid boundary—is the nexus of colloidal stability, electrokinetic transport, and interfacial electrochemistry. Its signature parameter, the ζ potential, determines the electrostatic effect that govern phenomena from protein adsorption to microfluidic pumping to the performance of energy storage devices. Yet for all its importance, the spatial distribution of ζ potential on a surface has remained essentially challenging.

Conventional streaming potential measurements yield only a spatial average across macroscopic electrode spans (*1-3*) (Fig. 1a, c). This averaging conceals any local heterogeneity, forcing researchers to adopt a simplifying assumption that has persisted for over a century: on a chemically uniform surface, the ζ potential is uniform. However, advanced techniques have begun to challenge this picture. Atomic force microscopy with colloidal probes offers improved spatial resolution but introduces invasive mechanical contact that perturbs the very interface under study (*4*). Vibrational sum-frequency generation spectroscopy has revealed that flow can alter surface charge density on specific calcium fluoride and fused silica substrates (*5, 6*)—but these measurements integrate over illumination spots and no reports provide full landscape of ζ potential on a surface.

Here we introduce the Fluorescence Photobleaching Electrochemistry Analyzer (FLEA), a non-invasive optical technique that maps local ζ potential with 180 nm lateral resolution. FLEA combines fluorescence photobleaching velocimeter with oscillating electroosmotic flow to probe the EDL at the shear plane without disturbing the interface (Fig. 1b). Using this tool, we examine a model system: pressure-driven laminar flow in a glass microchannel. The glass surface is chemically homogeneous, its silanol chemistry well-characterized—the ideal testbed for the uniformity hypothesis.

Our results overturn the assumption. We find that ζ potential varies not only along the flow direction—a phenomenon recently reported by others (*5, 6*)—but, more significantly, across the channel width, forming a nearly symmetrical two-dimensional (2D) pattern. This V-shaped distribution is not an artifact; it emerges systematically with flow and depends sensitively on physicochemical conditions. More importantly, this 2D electrochemical heterogeneity generates a non-conservative electric body force (EBF) that injects vorticity into the adjacent fluid, spontaneously creating streamwise vortices in boundary layers. This discovery reveals a new electrokinetic pathway to streamwise vorticity and establishes a fundamental connection between surface electrostatics and hydrodynamic structure.

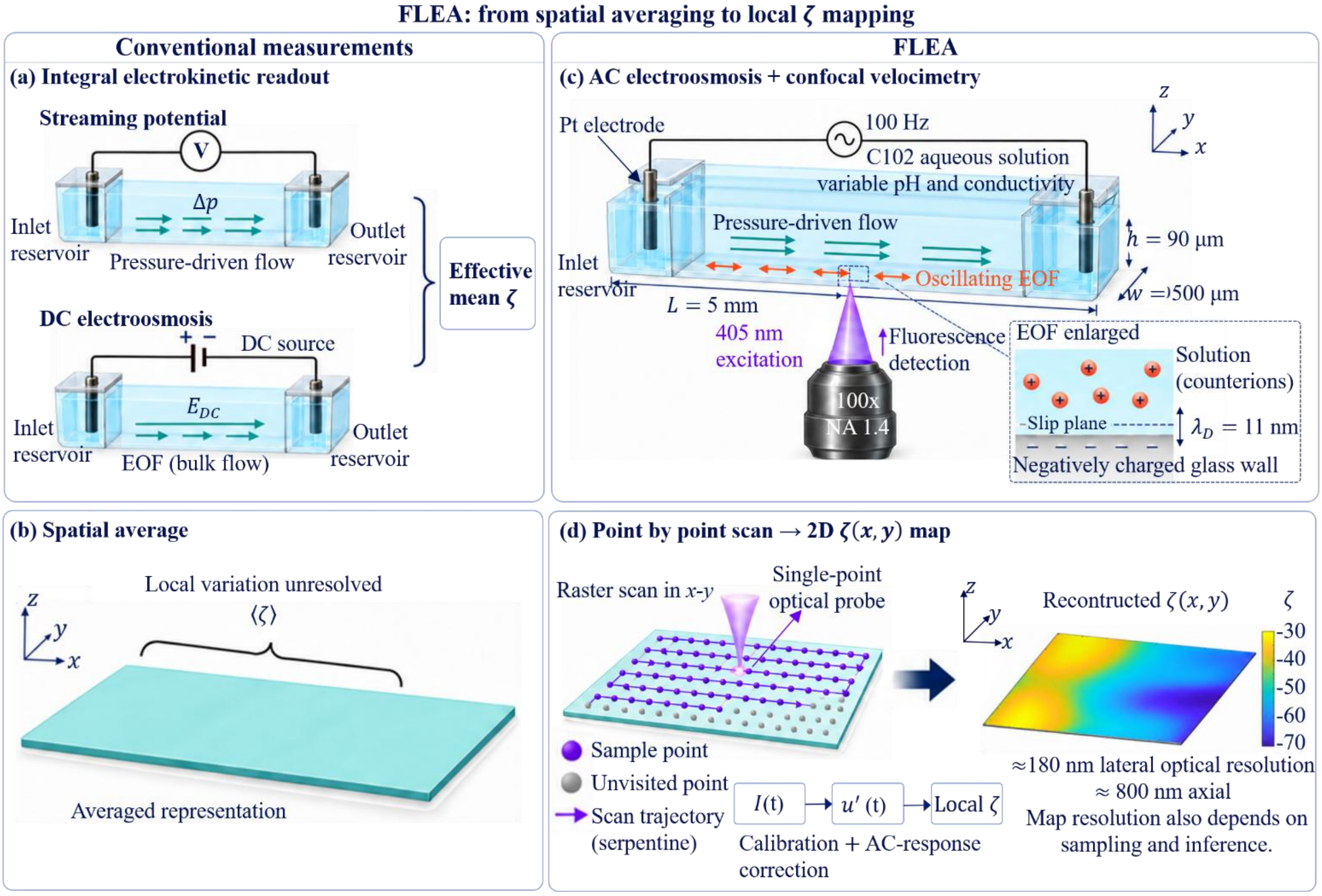

**Fig. 1 Schematic diagram of FLEA technique.** Conventional ways, e.g. streaming potential and DC electroosmosis, function using electroosmotic flow driven by direct current electric field (a), and reflects a spatially averaged $\zeta$ potential on the solid-liquid interface (b). In contrast, FLEA integrates oscillating electroosmotic flow (EOF) and velocimeter (c) to realize single-point measurement on local $\zeta$ potential distribution (d). In this investigation, we used FLEA to measure the $\zeta$ potential distribution in a microchannel with pressure-driven laminar flow ($x$-direction). The microchannel is $L = 5$ mm long, $w = 500$ μm wide and $h = 90$ μm high. Two platinum electrodes are placed in the inlet and outlet to generate oscillating EOF driven by AC electric field. The working fluid is Coumarin 102 (C102) aqueous solution of different electric conductivity and pH values. The thickness of EDL can be evaluated by Debye length which is $\lambda = 11$ nm.

## 2. Results

### 2.1 FLEA enables local, non-invasive and high-resolution $\zeta$ potential measurement

The FLEA system (Fig. S4a) integrates a home-built confocal microscope with a 405-nm CW laser, 100× oil-immersion objective (NA 1.4), and a nano-positioning stage. The spatial resolution reaches 180 nm laterally and 800 nm axially—sufficient to resolve variations across the microchannel cross-section. With a velocity sensitivity low to submicron to micrometers per second (*7*), it enables accurate measurement of the minute oscillatory flows induced by the probing AC electric field.

The measurement protocol is as follows (details in SI, section 3): A steady, pressure-driven laminar flow ($Re = \langle U \rangle D \rho_f / \eta = 1.13$) is established in the microchannel ($L = 5$ mm, $w = 500$ μm, $h = 90$ μm). This flow, we will show, patterns the surface potential. To read out the local $\zeta$ potential without perturbing this steady state, we superimpose a small-amplitude high-frequency AC electric field ($f_f = 100$ Hz). This generates a linear oscillating electroosmotic flow (OEOF) whose velocity fluctuation ($u_{rms}$) at the EDL shear plane is directly proportional to the local $\zeta$ potential via the Helmholtz–Smoluchowski relation (*8, 9*)

$$\zeta = \frac{\eta}{\varepsilon_0 \varepsilon_r} \frac{u_{rms}}{E_{rms}} \tag{1}$$

where $\varepsilon_r$ and $\varepsilon_0$ are the relative and vacuum permittivity, $\eta$ is dynamic viscosity of solution, $E_{rms}$ is the root-mean-square of AC electric field.

We measure $u_{rms}$ using fluorescence photobleaching (*10-16*): the fluorescent intensity of Coumarin 102 tracer molecules decreases with increasing velocity due to reduced residence time in the laser spot. A calibration curve (Fig. S2) converts fluorescence intensity to local velocity. Importantly, we operate at sufficiently high frequency (100 Hz) to avoid electrode polarization, electrolysis, and nonlinearities—ensuring that the measured $u_{rms}$ faithfully reports the local $\zeta$ potential.

By scanning the laser spot across the microchannel floor, we construct maps of local $\zeta$ potential with unprecedented spatial detail. The chemical environment (pH 5–9, conductivity 55–100 µS/cm) is controlled using NaOH, HCl, and phosphate-buffered saline (PBS) (see section 4 of SI). This systematic approach allows us to interrogate the spatial structure of $\zeta$ potential under well-defined conditions.

### 2.2 **A 2D $\zeta$ potential landscape emerges on a chemically uniform surface**

We first focus on the coarse $\zeta$ potential distribution across the microchannel. Fig. 2 presents the complete spatial maps where two intriguing features emerge.

First, the $\zeta$ potential varies along the flow direction (Fig. 2a–c), becoming increasingly negative downstream by up to 15 mV over 450 µm—consistent with the flow-induced charge redistribution reported by Ober et al. (*6*) and Werkhoven et al. (*17*). This effect scales with pH and conductivity: higher pH (up to 9) and higher conductivity amplify the downstream gradient.

As the pH value is increased from 5 to 7, an increasingly negative $\zeta$ potential can be observed, accompanied by the decreased $H^+$ concentration. The accumulation of negative charge on the solid-liquid interface is promoted, primarily due to the progressive deprotonation of surface silanol groups. At the silica surface, silanol sites exist in an acid–base equilibrium

$$\equiv \text{Si–OH} \rightleftharpoons \equiv \text{Si–O}^- + \text{H}^+ \qquad (2)$$

At pH=5, the relatively high proton activity shifts this equilibrium toward the protonated, neutral $\equiv$ Si–OH form, resulting in a lower density of negative charge on the interface, and hence a less negative $\zeta$ potential. As the pH is raised towards 7, the equilibrium shifts to the right, increasing the concentration of negatively charged silanolate groups ($\equiv$ Si–O$^-$) at the interface. This higher density of negative charge on the interface leads to a more negative $\zeta$ potential.

When the pH value is further increased, first, $\zeta$ potential is increased at pH=8 and then more negative again at pH=9. The non-monotonic variation of $\zeta$ potential could arise from the competition between pH-dependent surface deprotonation and the increasing ionic strength caused by NaOH addition. Although the pH elevation tends to make the $\zeta$ potential more negative, the added NaOH for adjusting pH value simultaneously introduces $Na^+$ ions, which increase the ionic strength of the solution. Elevated ionic strength compresses the EDL, thereby reducing the magnitude of the $\zeta$ potential even if the surface charge density continues to rise. This observation is consistent with the study of Scales et al (*18*), who demonstrate that higher ionic strength reduces the magnitude of the $\zeta$ potential at a given pH. In addition, $Na^+$ can screen or partially associate with the negatively charged surface groups, further attenuating the apparent $\zeta$ potential (*19*). We note that at pH 7, the gradient vanishes—a point we attribute to the balance between convective ion transport and proton-exchange kinetics.

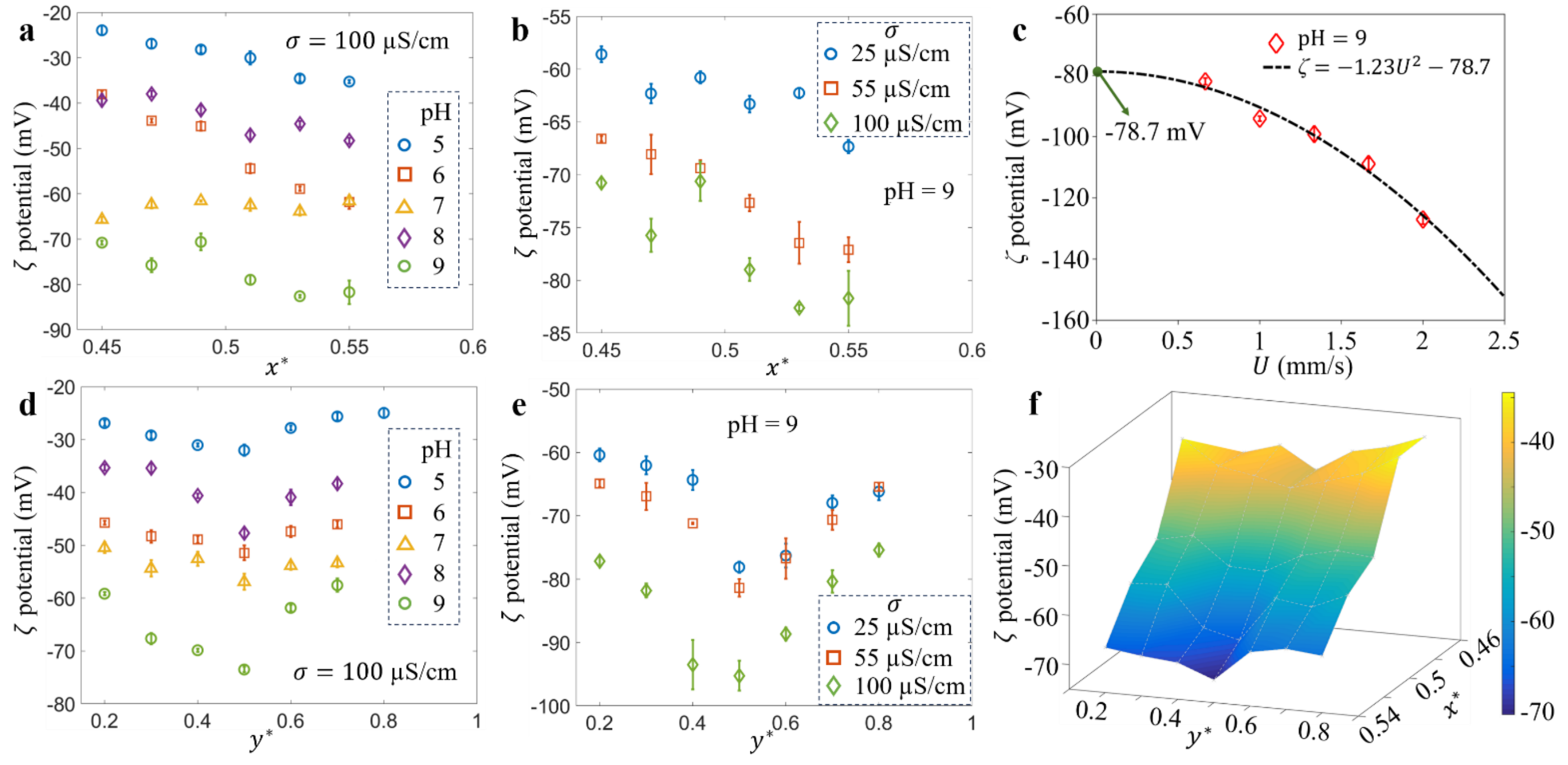

**Figure 2 | Distribution of $\zeta$ potential induced by external flow. a, b,** $\zeta$ potential distribution along the flow direction at $y^* = y/w = 0.5$, i.e., the centerline. All the $\zeta{\sim}x^*$ plots declined towards downstream indicates a decreasing of $\zeta$ potential along streamwise direction. **d, e,** $\zeta$ potential distribution along spanwise direction at $x^* = x/L = 0.5$. The minimum $\zeta$ potential locates at the center of microchannel and $\zeta$ potential is approximately symmetric in spanwise distribution. **a, d,** $\sigma = 100$ µS/cm, different pH value. **b, e,** pH=9, different conductivity. **c,** $\zeta$ potential vs flow velocity $U$ (where $U = |\vec{u}|$) measured at the centerline of microchannel, where $\sigma = 100$ µS/cm and pH=9. $\zeta$ potential is always more negative at higher $U$. **f,** 2D distribution of $\zeta$ potential induced by flow on the chemically uniform surface, where $\sigma = 100$ µS/cm and pH=9.

Second and also the unexpected observation, the $\zeta$ potential exhibits a pronounced spanwise variation (Fig. 2d–e). At any fixed streamwise position, the potential is most negative at the channel center and rises approximate symmetry toward the side walls. The profile is V-shaped, analog the parabolic velocity distribution of the pressure-driven flow. At the centerline, where the flow velocity is the highest, $\zeta$ reaches approximately –82.6 mV, while near the walls, it is less negative by 10–15 mV. This spanwise gradient is not a minor perturbation—it constitutes a genuine 2D landscape of $\zeta$ potential (Fig. 2f).

The spanwise distribution depends systematically on experimental conditions. Higher pH amplifies the center-to-wall difference (Fig. 2d), while higher conductivity both shifts the entire curve downward and steepens the gradient (Fig. 2e). Most significantly, the spanwise pattern scales with flow velocity (Fig. 2c): faster flow produces more negative $\zeta$. This establishes unequivocally that the flow is a crucial agent patterning the surface.

These results reveal a fundamental fact—when exposed to even laminar flow, a chemically homogeneous surface spontaneously acquires a 2D $\zeta$ potential distribution. The spanwise variation is particularly significant because it cannot be explained by any known mechanism—it represents a genuinely new phenomenon.

### 2.3 The 2D $\zeta$ potential generates a rotational electric body force

Despite the crucial influence of 2D $\zeta$ potential distribution on electrochemistry and the related applications, we want to ask what are the consequences of this 2D electrochemical patterning towards hydrodynamics. To address this, we turn to the Poisson–Boltzmann theory and the flow dynamic equations with EBF.

The 2D $\zeta$ potential map $\zeta(x,y)$ determines the electric potential field $\varphi(x,y,z)$ throughout the EDL via

$$\varphi(x,y,z)=\frac{4k_BT}{ez_i}\tanh^{-1}\left[\tanh\left(\frac{ez_i\zeta}{4k_BT}\right)\exp\left(-\frac{z}{\lambda}\right)\right] \qquad (3)$$

Here $\lambda = 11$ nm is the Debye length. The resulting EBF $\vec{F_e}=\rho_e\vec{E}=\varepsilon\nabla^2\varphi\nabla\varphi$ is predictable to be three-dimensional (3D) and non-uniform. These have been validated in Fig. 3a. $\vec{F_e}$ is primarily directed toward the wall (negative $z$-direction) and intensifies downstream. Its magnitude varies across the span, being largest at the center where the $\zeta$ gradient is steepest. This non-uniform, rotational force cannot be balanced by the pressure gradient (Fig. S5) alone. Thus, a critical insight emerges when we examine the curl of this force. As well-known, for a conservative force field, the curl is zero, i.e. irrotational. But here, the force acquires a non-zero curl (see SI, section 1 for full derivation), with

$$\vec{T}=\frac{1}{\rho_f}\left(\nabla\times\vec{F_e}\right)=\frac{\varepsilon}{\rho_f}\nabla^2(\nabla\varphi)\times\nabla\varphi\neq 0 \qquad (4)$$

This is a pivotal result. The curl of $\vec{F_e}$ provides a source term in the steady vorticity ($\vec{\Omega}=\nabla\times\vec{u}$) transport equation

$$(\vec{u}\cdot\nabla)\vec{\Omega}-(\vec{\Omega}\cdot\nabla)\vec{u}=\nu\nabla^2\vec{\Omega}+\vec{T} \qquad (5)$$

**The 2D $\zeta$ potential distribution thus directly injects vorticity into the fluid.** This is impossible if the potential is uniform or varies only along the flow direction, since $\vec{T}=0$ according to Eq. (S12). It is the **spanwise variation**—the new discovery enabled by FLEA—that makes the electric force rotational and generates vorticity.

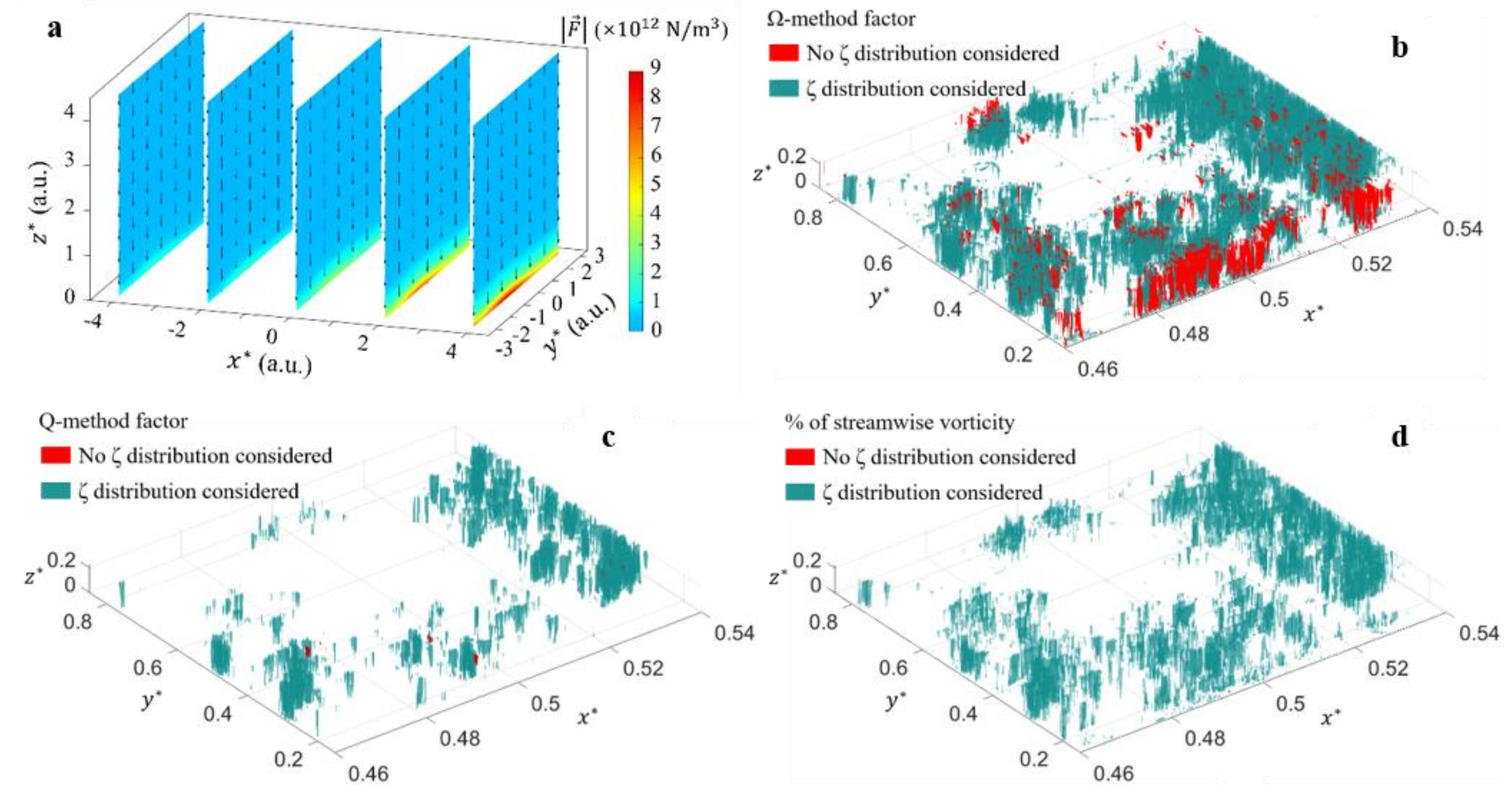

**Figure 3 | a,** EBF $\vec{F_e}$ generated according to the 2D distribution of $\zeta$ potential induced by laminar basic flow. $\vec{F_e}$ has a larger z-directional components and towards the bottom wall. Its magnitude increases along streamwise direction, and exhibits asymmetry in spanwise. **b,** the vortices distinguished by Ω method with a threshold of 0.52. **c, strong vortices region** distinguished by Q method with a threshold of $10^4$ s$^{-1}$. **d,** streamwise vortices where $|\Omega_x|/|\vec{\Omega}|\geq 0.7$. All the results are simulated by COMSOL Multiphysics software.

2.4 Streamwise vortices emerge from the electrokinetic feedback loop

To examine the flow structures implied by this rotational forcing, we performed three-dimensional simulations in COMSOL Multiphysics (Fig. 3). The computational domain (450 × 350 × 90 μm³) covers the region in which $\zeta$ was measured. The measured $\zeta(x, y)$ at $\sigma = 100$ μS/cm and pH 9 was fitted with a 2D polynomial surface (Eq. (S14), with $R^2 = 0.96$) and imposed through Eq. (3). A structured hexahedral mesh ( $\sim 1.5 \times 10^7$ elements) resolves the EDL with 1.33-nm spacing normal to the wall and 0.53–1.6 μm spacing parallel to it. A control simulation with identical geometry and flow but without the EBF serves as the reference.

With three complementary criteria (Fig. 3b–d), we reveal a dramatic transformation of the flow field. The Ω-method (threshold 0.52 (20)) is a normalized measure and is therefore sensitive even to weak vortices. We compare two cases: the control simulation without the influence of $\zeta$ potential (the traditional assumption) and the experimental case with the measured 2D $\zeta$ potential. The difference is striking (Fig. 3b). **In the former, virtually few vortices (red) are detected. While in the latter, numerous vortices (green) emerge across the computational domain.** Some of the vortices are so small even under 1 μm, which is much smaller than Kolmogorov scale in regular fluids, and the vortices in superconductors (*21*) and quantum fluids (*22*).

The Q-criterion (*23, 24*) with a large threshold at $10^4$ $s^{-1}$ measures the absolute strength of rotation relative to strain and separates the two cases more clearly: vortical regions are nearly absent in the control but widespread when the 2D $\zeta$ landscape is included (Fig. 3c). These are not numerical artifacts but genuine rotational structures driven by the EBF.

**Most remarkably, a significant fraction of these vortices is oriented in the streamwise direction.** We identify them by requiring that the streamwise vorticity component exceeds 70% of the total ($|\Omega_x|/|\vec{\Omega}| \geq 0.7$). The resulting streamwise vortex tubes extend along the channel (Fig. 3d). This is deeply counterintuitive: **laminar flow over a flat, chemically uniform wall, at $Re \approx 1$, spontaneously generates streamwise vortices.** The mechanism is purely electrokinetic, arising from the feedback between hydrodynamic convection and interfacial electrochemistry. The 2D $\zeta$ potential, measured by FLEA, is the key that unlocks this new route to vorticity.

## 3. Discussion and conclusions

Our results establish three interconnected discoveries that reshape our understanding of interfacial electrokinetics.

First, FLEA provides a transformative capability for surface characterization, with non-invasive 180 nm spatial resolution. This capability opens immediate applications in broad areas, for instance, characterizing novel surface coatings and functionalized interfaces, developing heterogeneous catalysts devices, designing biosensors for selective protein adsorption, and fabricating metamaterial surfaces with engineered charge distributions etc. Unlike streaming potential measurements, which average over macroscopic areas, FLEA reveals the hidden spatial structure that governs local electrokinetic phenomena.

Second, we have discovered that flow patterns a chemically uniform surface into a 2D $\zeta$ potential landscape. The spanwise V-shaped distribution—most negative at the center, less negative near the walls—is particularly significant and unpredicted. It reflects the coupling between convective ion transport and surface chemistry, generating spatial heterogeneity. At higher velocities (center of channel), the convective flux of charged species is greater, enhancing the accumulation of negative charges (deprotonated silanols) at the interface. Near the walls, where velocity approaches zero, this convective enhancement is minimal.

We acknowledge the investigations (*6, 17*) that reveal flow-along variation of $\zeta$ potential. The novelty of our work is the spanwise heterogeneity and the complete 2D map—enabled exclusively by FLEA's spatial resolution.

Third, and most significantly, the 2D $\zeta$ potential landscape generates streamwise vortices at the small scales. This discovery reveals an entirely new physical mechanism for vortex formation at interfaces: electrokinetic feedback driven by spatially varying surface potential. It is fundamentally different from the classical routes to streamwise vorticity: shear-layer instabilities, roughness-induced transitions, or secondary flows in curved geometries. It operates at $Re \approx 1$ —far below any hydrodynamic instability threshold. It relies solely on the electrochemical properties of the interface and their coupling to convection. This raises the intriguing possibility that electrokinetic feedback in the boundary layer may provide the initial "seeds" that, under turbulent conditions, develop into hairpin vortices through classical vortex stretching and amplification mechanisms. The connection, if confirmed, would establish a fundamental link between interfacial electrochemistry and the dynamics of wall-bounded turbulence—a link that has never before been considered.

We hope these findings can also shed light on current observations in interfacial sciences. Overall, this breakthrough has the capacity to greatly advance the development of novel and complex chemical and biomedical materials for broad applications.

**Acknowledgement** The investigation is supported by National Natural Science Foundation No. 51927804.

# Supplementary Information

# Flow induced two-dimensional ζ potential on chemically uniform solid-liquid interface: A possible electrokinetic origin of streamwise vortices

**Shuangshuang Meng, [1,#] Yu Han, [1,#], Xiaoqiang Feng, [1] Yueqiang Zhu, [1] Jin'an Pang, [1] Duyang Zang, [2] Yu Chen, [1] Ruihong Zhang, [1] Chen Zhang, [1] Ce Zhang, [1,*] Kaige Wang, [1,*] Jintao Bai, [1] and Wei Zhao [1,*]**

[1] State Key Laboratory of Photon-Technology in Western China Energy, International Collaborative Center on Photoelectric Technology and Nano Functional Materials, Institute of Photonics & Photon Technology, Northwest University, Xi'an 710127, China

[2] School of Physical Science and Technology, Northwestern Polytechnical University, Xi'an 710129, P. R. China

* Corresponding author: zhangce@nwu.edu.cn; wangkg@nwu.edu.cn; zwbayern@nwu.edu.cn

\# These authors contribute equally to this investigation

**This PDF file includes:**

## S1. Theory

According to Poisson–Boltzmann theory, the electric potential in the electric double layer at a distance $z$ from the interface can be expressed as:

$$\varphi(x,y,z)=\frac{4k_BT}{ez_i}\tanh^{-1}\left[\tanh\left(\frac{ez_i\zeta}{4k_BT}\right)\exp\left(-\frac{z}{\lambda}\right)\right] \tag{S1}$$

where $\zeta=\zeta(x,y)$ is $\zeta$ potential in 2D, $\lambda=\sqrt{\varepsilon_0\varepsilon_r k_BT/2e^2N_A\sum c_iz_i^2}$ is Debye length, $k_B$ is Boltzmann constant, $T$ is temperature, $e$ is elementary charge, $\varepsilon_0$ and $\varepsilon_r$ are vacuum and relative permittivity respectively, $N_A$ is Avogadro constant, $c_i$ and $z_i$ are the ionic concentrating and valence respectively.

The steady-state Navier–Stokes equation with electric body force ($\overrightarrow{F_e}$) is

$$\vec{u}\cdot\nabla\vec{u}=-\frac{1}{\rho_f}\nabla P+\nu\nabla^2\vec{u}+\frac{1}{\rho_f}\overrightarrow{F_e} \tag{S2}$$

where $\rho_f$ is fluid density, $P$ is the pressure on fluid, $\nu$ is kinematic viscosity and $\vec{u}=u_i\widehat{x_i}$ is the instant velocity vector ($u_i$ is the velocity components in $\widehat{x_i}$ direction, $\widehat{x_1}$, $\widehat{x_2}$, and $\widehat{x_3}$ are corresponding unit vectors in streamwise, spanwise and vertical directions respectively). In the following section, we use $x_1$, $x_2$, and $x_3$ to represent $x$, $y$, and $z$ for convenience. Sum convention has been applied here. $\overrightarrow{F_e}=\rho_e\vec{E}$, where $\rho_e$ is net charge density and $\vec{E}$ is electric field. These quantities can be expressed as

$$\rho_e=-\varepsilon\nabla^2\varphi \tag{S3}$$

$$\vec{E}=-\nabla\varphi \tag{S4}$$

Assuming incompressible flow, we take the curl on both sides of Eq. (S2), and get

$$(\vec{u}\cdot\nabla)\vec{\Omega}-(\vec{\Omega}\cdot\nabla)\vec{u}=\nu\nabla^2\vec{\Omega}+\vec{T} \tag{S5}$$

where $\vec{T}=(\nabla\times\overrightarrow{F_e})/\rho_f$. According to Maxwell's equations,

$$\nabla\times\overrightarrow{F_e}=\nabla\rho_e\times\vec{E}+\rho_e\nabla\times\vec{E}=\nabla\rho_e\times\vec{E}-\rho_e\frac{\partial\vec{B}}{\partial t} \tag{S6}$$

Although the streaming current can generate a magnetic field, $\vec{B}$ is steady, and thus $\partial\vec{B}/\partial t=0$. Eq. (S6) therefore reduces to

$$\nabla\times\overrightarrow{F_e}=\nabla\rho_e\times\vec{E} \tag{S7}$$

For constant and quasi-uniform $\varepsilon$, $\nabla\varepsilon=0$. Combining with Eqs. (S3) and (S4), Eq. (S7) becomes

$$\nabla\times\overrightarrow{F_e}=\varepsilon\nabla^2(\nabla\varphi)\times\nabla\varphi \tag{S8}$$

Let $f_i=\partial f/\partial x_i$, $f_{ij}=\partial^2 f/\partial x_i\partial x_j$, $f_{ijk}=\partial^3 f/\partial x_i\partial x_j\partial x_k$, with $f$ representing either $\zeta$ or $\varphi$, then

$$\nabla\varphi=\frac{BN\zeta_1}{M}\widehat{x_1}+\frac{BN\zeta_2}{M}\widehat{x_2}+\frac{\psi AB}{\lambda M}\widehat{x_3} \tag{S9}$$

$$\nabla^2\varphi=\varphi_{ii}=\frac{BN\zeta_{11}}{M}-\frac{2ABN\zeta_1^2}{\psi M}+\frac{2AB^3N^2\zeta_1^2}{\psi M^2}+\frac{BN\zeta_{22}}{M}-\frac{2ABN\zeta_2^2}{\psi M}+\frac{2AB^3N^2\zeta_2^2}{\psi M^2}+\frac{2\psi A^3B^3}{\lambda^2M^2}-\frac{\psi AB}{\lambda^2 M} \tag{S10}$$

where $A(x_1,x_2)=\tanh[\zeta(x_1,x_2)/\psi]$, $B(x_3)=\exp(-x_3/\lambda)$, $M=A^2B^2-1$, $N=A^2-1$ and $\psi=4k_BT/ez_i$. Thus,

$$\overrightarrow{F_e}=\varepsilon\nabla^2\varphi\nabla\varphi=\frac{\varepsilon B^2}{M^2}\left(N\zeta_{11}-\frac{2AN\zeta_1^2}{\psi}+\frac{2AB^2N^2\zeta_1^2}{\psi M}+N\zeta_{22}-\frac{2AN\zeta_2^2}{\psi}+\frac{2AB^2N^2\zeta_2^2}{\psi M}+\frac{2\psi A^3B^2}{\lambda^2M}-\frac{\psi A}{\lambda^2}\right)\left(N\zeta_1\widehat{x_1}+N\zeta_2\widehat{x_2}+\frac{\psi A}{\lambda}\widehat{x_3}\right) \tag{S11}$$

Let $\epsilon_{ikk'}$ being Levi-Civita tensor, then the $\vec{T}$ vector according to Eq. (S5) near the insulated solid-liquid interface becomes

$$\vec{T}=\frac{\varepsilon}{\rho_f}\nabla^2(\nabla\varphi)\times\nabla\varphi=\frac{\varepsilon}{\rho_f}(\varphi_i)_{jj}\varphi_k\widehat{x_{k'}}\epsilon_{ikk'} \tag{S12}$$

with

$$T_1=\frac{\varepsilon B^2N}{\rho_f\lambda M^2}\left\{\frac{4A}{\psi}\left(A^2+\frac{B^2N^2}{M}-\frac{2A^2B^2N}{M}\right)\zeta_2(\zeta_1^2+\zeta_2^2)+\left[(N+\frac{4A^2B^2N}{M}-6A^2)\zeta_2\zeta_{22}+(N-2A^2)\zeta_2\zeta_{11}+4A^2\left(\frac{B^2N}{M}-1\right)\zeta_1\zeta_{12}\right]+\psi A(\zeta_{112}+\zeta_{222})\right\}$$

$$T_2=-\frac{\varepsilon B^2N}{\rho_f\lambda M^2}\left\{\frac{4A}{\psi}\left(A^2+\frac{B^2N^2}{M}-\frac{2A^2B^2N}{M}\right)\zeta_1(\zeta_2^2+\zeta_1^2)+\left[(N+\frac{4A^2B^2N}{M}-6A^2)\zeta_1\zeta_{11}+(N-2A^2)\zeta_1\zeta_{22}+4A^2\left(\frac{B^2N}{M}-1\right)\zeta_2\zeta_{12}\right]+\psi A(\zeta_{122}+\zeta_{111})\right\}$$

$$T_3=\frac{\varepsilon B^2N^2}{\rho_f M^2}\left\{\frac{4A}{\psi}\left(\frac{B^2N}{M}-1\right)[\zeta_{12}(\zeta_2^2-\zeta_1^2)+\zeta_1\zeta_2(\zeta_{11}-\zeta_{22})]+[\zeta_2(\zeta_{111}+\zeta_{122})-\zeta_1(\zeta_{112}+\zeta_{222})]\right\}$$

being the components of $\vec{T}$ in three directions respectively. $\zeta_i\zeta_j\zeta_k$, $\zeta_i\zeta_{jk}$ and $\zeta_{ijk}$ have 4, 6 and 4 combinations respectively, as $i,j,k$ are either 1 or 2. Since all $\zeta_i\zeta_j\zeta_k$, $\zeta_i\zeta_{jk}$ and $\zeta_{ijk}$ quantities are nonzero, according to Eq. (S12), we have

$$\vec{T}\neq 0 \tag{S13}$$

Thus, we have theoretically proven that vortical flow can be induced in the vicinity of the EDL at an insulated solid–liquid interface.

## S2. Numerical Simulations

S2.1 Modeling and meshing

We used COMSOL Multiphysics to solve Eqs. (S1)–(S4) and simulate the velocity, vorticity, helicity, and electric body force fields. In the numerical model, the computation was performed in a rectangular domain of 450 μm long, 350 μm wide, and 90 μm high, to be consistent with the measurement region of $\zeta$ potential in experiments. The solution in the measurement of $\zeta$ potential was Coumarin 102 fluorescent solution. HCl, NaOH and phosphate-buffered saline (PBS) were added to adjust the ion concentration and pH value of the solution. The simulation was conducted at $\sigma = 100$ μS/cm and pH=9, with the Debye length is $\lambda = 11$ nm.

To handle the large difference (approximately four to five orders of magnitude) in dimensions between the electric double layer and the microchannel in the $z$ direction, the computational domain was divided into three layers (Fig. S1a). This approach reduces the computational burden of resolving the flow in 3D across such a wide range of length scales.

The three layers consist of two thin-layer regions near the solid-liquid interface (Fig. S1b), and a bulk region located at the center of the channel (Fig. S1c). The height of the thin-layer regions, denoted as $h_s$, is set to 50 nm. To compromise between resolution and computation cost, a condensed mesh is used in the domain near the measurement positions of the $\zeta$ potential in the *x-y* plane. During the grid construction process, a mapping operation is employed in the *x-y* plane, while a sweeping operation is carried out upwards from the bottom to the top in the *z* direction. When sweeping through the two thin-layer regions, ultrafine grids are applied to ensure a nanoscale resolution in the *z* direction.

Hexahedral mesh elements are used in both the bulk region and the thin-layer regions. In the bulk region, the grid has a fixed size of 17.98 μm in the *z* direction. In the thin-layer region, the grid sizes vary, with a minimum size of 1.33 nm and a maximum size of 2.66 nm in the *z* direction. In the *xy* plane, during the mapping and sweeping operations, the grid sizes remain the same in both the bulk region and the thin-layer regions. In the *x* direction, the minimum and maximum grid sizes are 0.53 μm and 1.61 μm, respectively, in both regions. In the y direction, the minimum and maximum grid sizes are 0.53 μm and 1.58 μm, respectively.

Figure S1 provides a schematic diagram of the model, showing its dimensions of 450 μm×350 μm×90 μm. Different regions are distinguished by the volume of the grid cells, with mirror symmetry between layers 1 and 3. The two thin layers are magnified 500 times in the $z$ direction for better visibility. The total number of grid cells in the model is 15,060,357. For detailed parameters and grid statistics of each region, refer to Extended Data Table 1.

S2.2 Physical model and boundary conditions

The basic flow is pressure-driven laminar flow, as both the experimental microchannel size and the bulk flow Reynolds number are very small. The effects of gravity and compressibility are neglected, and the flow is assumed to be steady state. An incompressible laminar flow model is selected for the study.

To investigate the impact of the non-uniform $\zeta$ potential on the flow field, an electrostatic model is incorporated into the simulation. To improve computational efficiency, custom variables and analytic functions are used to calculate the electric body force. This involves fitting the experimentally measured 2D distribution of $\zeta$ potential to a polynomial two-dimensional function (by Matlab 2025b), which has three orders in $x$ direction and five orders in $y$ direction, as

$$\begin{aligned}\zeta(x', y') = &-53.78 - 0.127x' + 0.04446y' - 4.602 \times 10^{-5}x'^2 \\ &+7.264 \times 10^{-5}x'y' + 8.819 \times 10^{-4}y'^2 + 1.517 \times 10^{-6}x'^3 \\ &+3.69 \times 10^{-7}x'^2y' + 5.095 \times 10^{-7}x'y'^2 - 6.669 \times 10^{-6}y'^3 \\ &-9.239 \times 10^{-10}x'^3y' - 2.044 \times 10^{-9}x'^2y'^2 - 3.157 \times 10^{-9}x'y'^3 \\ &-2.625 \times 10^{-8}y'^4 - 2.136 \times 10^{-12}x'^3y'^2 + 3.67 \times 10^{-12}x'^2y'^3 \\ &-2.069 \times 10^{-11}x'y'^4 + 2.029 \times 10^{-10}y'^5 \end{aligned} \quad (S14)$$

where $x' = x - x_c$ and $y' = y - y_c$ are the coordinates relative to the center ($x_c = 2.5$ mm, $y_c = 250$ μm) of microchannel, which is also the center of $\zeta$ potential measurement. They are in dimensions of μm. $\zeta$ has a dimension of mV. Then, the electric body force is then determined using the Poisson–Boltzmann theory and the Poisson equation.

Different boundary conditions are applied in the model. At the microchannel inlet, a fully developed flow condition with an average velocity of $U_{av} = 6.78 \times 10^{-5}$ m/s is specified, corresponding to a flow rate $Q = 20$ μL/min. At the exit, a static pressure of $P_0 = 0$ Pa is set, along with a reflux inhibition condition to prevent backflow. No-slip boundary conditions are applied to the upper and bottom walls, as well as the side walls, consistent with the symmetry of the geometric and physical model.

## S3. Principle of $\zeta$ potential measurement

The Fluorescence Photobleaching Electrochemistry Analyzer (FLEA) system combines velocity measurement by laser-induced fluorescence photobleaching, AC oscillatory electroosmotic flow, and numerical analysis methods for local $\zeta$ potential measurement at the insulated solid–liquid interface.

**Extended Data Table.1** Detailed information of grid sizes in different computation regions.

| Regions | Grid sizes | x direction | y direction | z direction |
|---|---|---|---|---|
| **Thin-layer regions (layer 1 and layer 3)** | Maximum grid size | 1.61 μm | 1.58 μm | 2.66 nm |
| | Minimum grid size | 0.53 μm | 0.53 μm | 1.33 nm |
| **Bulk region (layer 2)** | Maximum grid size | 1.61 μm | 1.58 μm | 17.98 μm |
| | Minimum grid size | 0.53 μm | 0.53 μm | 17.98 μm |
| **Overall** | Grid number | 401 | 351 | 107 |

S3.1 Velocity measurement by laser induced fluorescence photobleaching anemometer

Flow velocity is determined using laser-induced fluorescence photobleaching. When a fluorescent dye at a specific concentration is exposed to a certain laser power, an increase in flow velocity reduces the residence time within the laser irradiation area, resulting in less photobleaching and higher fluorescence intensity. The velocity can be obtained from a calibration curve that correlates fluorescence intensity ($I_f$) with local flow velocity, as illustrated in Fig. S2. For further details on this technique, please refer to our previous studies (*11-15, 25-27*).

S3.2 Principle of measuring $\zeta$ potential by FLEA

Under steady-state basic flow, the velocity fluctuation ($u_{rms}$) of the oscillating electroosmotic flow at the shear plane of the EDL (as shown in Fig. 1) is measured to calculate the $\zeta$ potential at the interface using the FLEA technique. The interface $\zeta$ potential is obtained by substituting $u_{rms}$ into the following equation

$$\zeta = k_s \frac{\eta}{\varepsilon_0 \varepsilon_r} \cdot \frac{u_{rms}}{E_{rms}} \tag{S15}$$

with

$$E_{rms} = \sqrt{\frac{1}{T}\int_0^T [E_A \sin(\omega t)]^2 dt} \tag{S16}$$

where, $k_s = sign[R_{uE}(0)]$ is a sign function with a value of $\pm 1$, $\eta$ is the dynamic viscosity, $\omega$ is the angular frequency, $T$ is the time, $E_{rms}$ is the electric field fluctuation, $E_A$ is the voltage amplitude, $R_{uE}(\tau)$ is the normalized (or normalized) mutual correlation function of the instantaneous velocity fluctuation $u(t)$ and the instantaneous electric field fluctuation $E(t)$, and $\Delta t$ is the time interval. $R_{uE}(\Delta t)$ can be calculated by

$$R_{uE}(\Delta t) = \frac{\sum_t u(t) \cdot E(t + \Delta t)}{\sqrt{\sum_t u^2(t) \sum_t E^2(t + \Delta t)}} \tag{S17}$$

To perform a FLEA measurement, first set the flow rate $Q$, electric field intensity $E$, and solution chemical parameters like pH value, solution concentration, and conductivity. Next, obtain velocity fluctuation data at various frequencies by measuring a set of $u_{rms} \sim f_f$ curves (Fig. S3). Here, we cannot directly measure $u_{rms}$ at low frequency where electrolysis effect is strong and can generate air bubble during a long-period measurement, although the measurement at low frequency is more accurate. Otherwise, it could affect the calculation of $\zeta$. Therefore, an indirect measurement of $u_{rms}$ has been conducted at a relatively higher $f_f = 100$ Hz. We use the average velocity fluctuation value ($\langle u_{rms} \rangle_r$) at low frequencies measured in a short period serves as reference value 1, while the velocity fluctuation value ($\langle u_{rms} \rangle_m$) at the higher frequency $f_f = 100$ Hz serves as reference value 2. Then, the proportional constant $k_c$ of $\langle u_{rms} \rangle_r$ measured in a short period and $\langle u_{rms} \rangle_m$ can be determined using Eq. (S18)

$$\langle u_{rms} \rangle_r = k_c \cdot \langle u_{rms} \rangle_m \tag{S18}$$

Then, according to Eq. (S17) again, by measuring $\langle u_{rms} \rangle_m$ at a higher frequency $f_f = 100$ Hz, the $\langle u_{rms} \rangle_r$ at low frequency can be determined. Finally, $\zeta$ can be calculated from Eq. (S15) through $\langle u_{rms} \rangle_r$.

## S4. Methods

S4.1 Experimental setup

The FLEA system uses a home-built confocal microscope, as depicted in Fig. S4a. The system offers a spatial resolution of 180 nm (lateral) and approximately 800 nm (axial), with a sampling rate of 1 kHz. A 15 mW, 405 nm CW laser (CW405, MDL-III-405-500, Changchun New Industries Optoelectronics Tech Co., China) serves as the light source. To improve the quality of the excitation beam, a spatial filter (SF, SFB-16DM, OptoSigma) is used. The excitation beam passes through a dichroic mirror (DM, Semrock), which transmits the 405 nm laser and reflects the 479 nm fluorescence. A Plan Apo 100× NA 1.4 oil-immersion objective (Olympus) focuses the excitation beam. Fluorescence is collected by the same objective, focused into an optical fiber, filtered, and detected by a photon counter (H7421-40, Hamamatsu). Precise positioning is achieved using a 3D micro-stage (M-521.DG, Physik Instrumente) and a nano-stage (P-562.3CD, Physik Instrumente).

S4.2 Microchip fabrication and solution preparation

The microchip is fabricated using a layer-by-layer process, with three layers: a cover layer, a middle layer, and a bottom layer, as illustrated in Fig. S4b. The cover layer is made of transparent acrylic plate with a thickness of 1.6 mm. The middle layer is made of transparent plastic plate with a thickness of 90 μm. The bottom layer is a cover glass with good UV transmittance and a thickness of 170 μm. The microchip contains a microchannel located in the middle layer. The microchannel has a length ($L$) of 5 mm, a width ($w$) of 500 μm, and a height ($h$) of 90 μm. Platinum electrodes are placed at both ends of the microchannel to provide an AC electric field.

Coumarin 102 (C102) fluorescent dye is used as markers. The excitation and fluorescent spectra of C102 are plotted in Fig. S4c. A C102 solution of 0.1 M is prepared by mixing C102 powder, 5% methanol, and 95% deionized water in proportion. The pH value of the solution is adjusted using NaOH and HCl buffer solution, and the electric conductivity is controlled using phosphate buffer saline (PBS).

S4.3 Experimental process

Prior to experimentation, the microchannel was rinsed with deionized water and C102 solution to ensure a clean environment for the measurements. A velocity calibration curve (Fig. S2) was then established, quantifying the monotonic relationship between fluorescence intensity and fluid flow velocity at the center of the microchannel, as shown in Fig. S2.

The $\zeta$ potential measurements were conducted at a basic flow rate ($Q$) of 20 μL/min, delivered by a syringe pump (PhD 2000, Harvard Apparatus). The applied AC electric field intensity ($E$) was set to $3 \times 10^3$ V/m, generated by an arbitrary function generator (AFG3102, Tektronix) and amplified by a high-voltage amplifier (PZD700A, Trek Inc.). The AC electric field was applied directly across the two platinum electrodes, as shown in Fig. S4b.

The root-mean-square velocity fluctuations were measured near the bottom wall at the interface where the fluctuating velocity within the electric double layer is maximal, as illustrated in Fig. 1. For each dataset, fluorescence was tracked for 30 seconds at a data acquisition rate of 1 kHz to compute the velocity fluctuation from Eq. (S18). To ensure reproducibility, each measurement was repeated five times under each experimental condition.

Initially, the $\langle u_{rms}\rangle_m$ was determined at $f_f = 100$ Hz. From these empirical observations, the rms velocity fluctuations $\langle u_{rms}\rangle_r$ at localized positions was calculated using Eq. (S18). By substituting these values into Eq. (S15), the localized $\zeta$ potential could hence be extracted.

Furthermore, $\langle u_{rms}\rangle_r$ was assessed under varying flow rates $Q$, whence $\zeta$ potentials corresponding to each flow rate were computed as before and subsequently fitted to acquire the $\zeta$ potential in the absence of flow using the following empirical relation

$$\zeta_{fit}(Q) = aQ^b + \zeta_{sta} \tag{S19}$$

Herein, $\zeta_{fit}$ denotes the fitted $\zeta$ potential, with $a$ and $b$ denoting the fitting parameters. In conditions where $Q$ approaches zero ($Q = 0$ μL/min), $\zeta_{fit}(0)$ converges to $\zeta_{sta}$, which is the quiescent $\zeta$ potential at the solid-liquid interface within a static flow domain. When changing conditions, e.g., positions, pH value and electric conductivity, the experimental processes above are repeated.

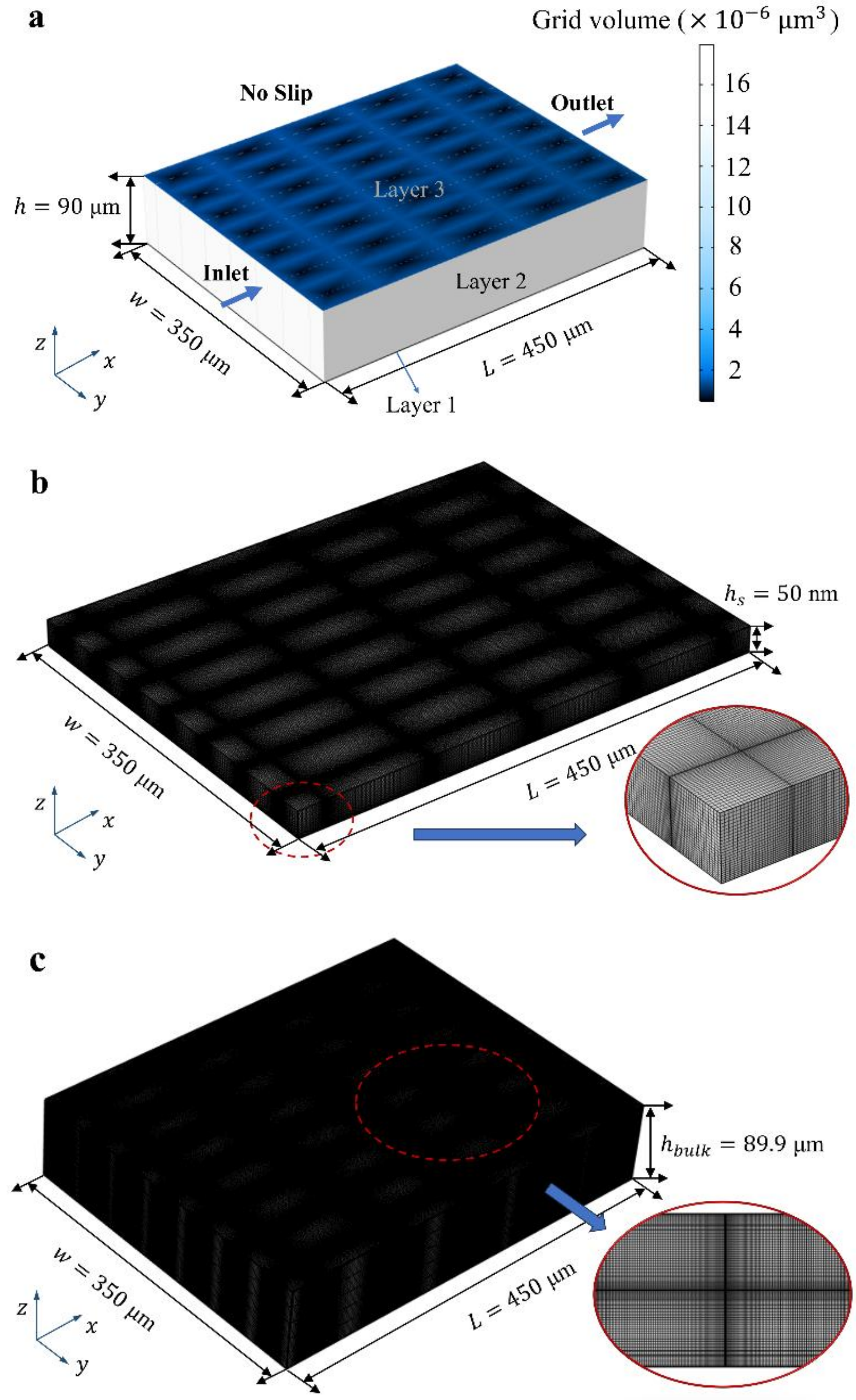

**Fig. S1 | Model diagram and grid distribution. a,** Schematic diagram of model and the distribution of grid volume. **b,** Grid of layer 1, i.e., thin-layer region around the solid-liquid interface, which is mirror symmetry with layer 3 (the schematic is scaled up 500 times in the z direction). **c,** Grid of layer 2, i.e., bulk region, which is away from the EDL, therefore, the grid can be coarsened appropriately. The inset shows the grid details of local area.

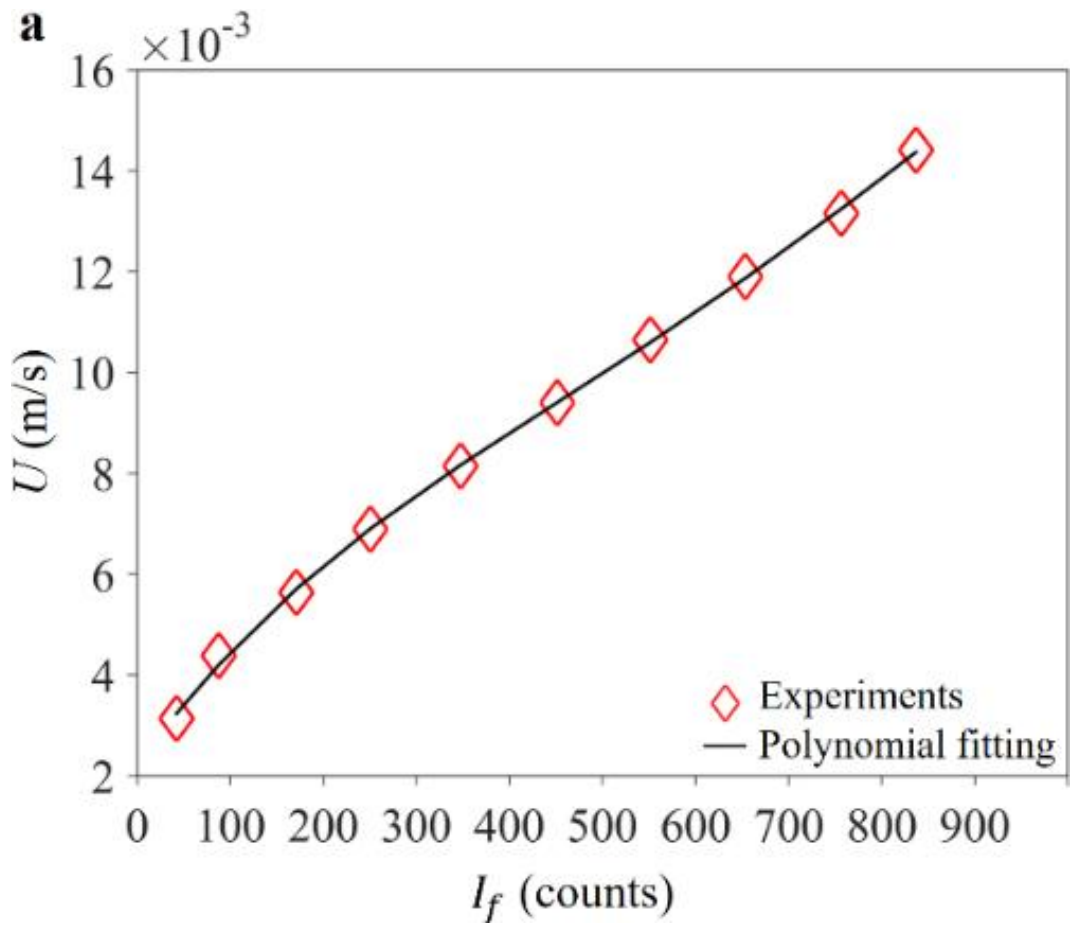

**Fig. S2 | Velocity calibration curve.** $I_f$ is fluorescent intensity, $U$ is flow velocity

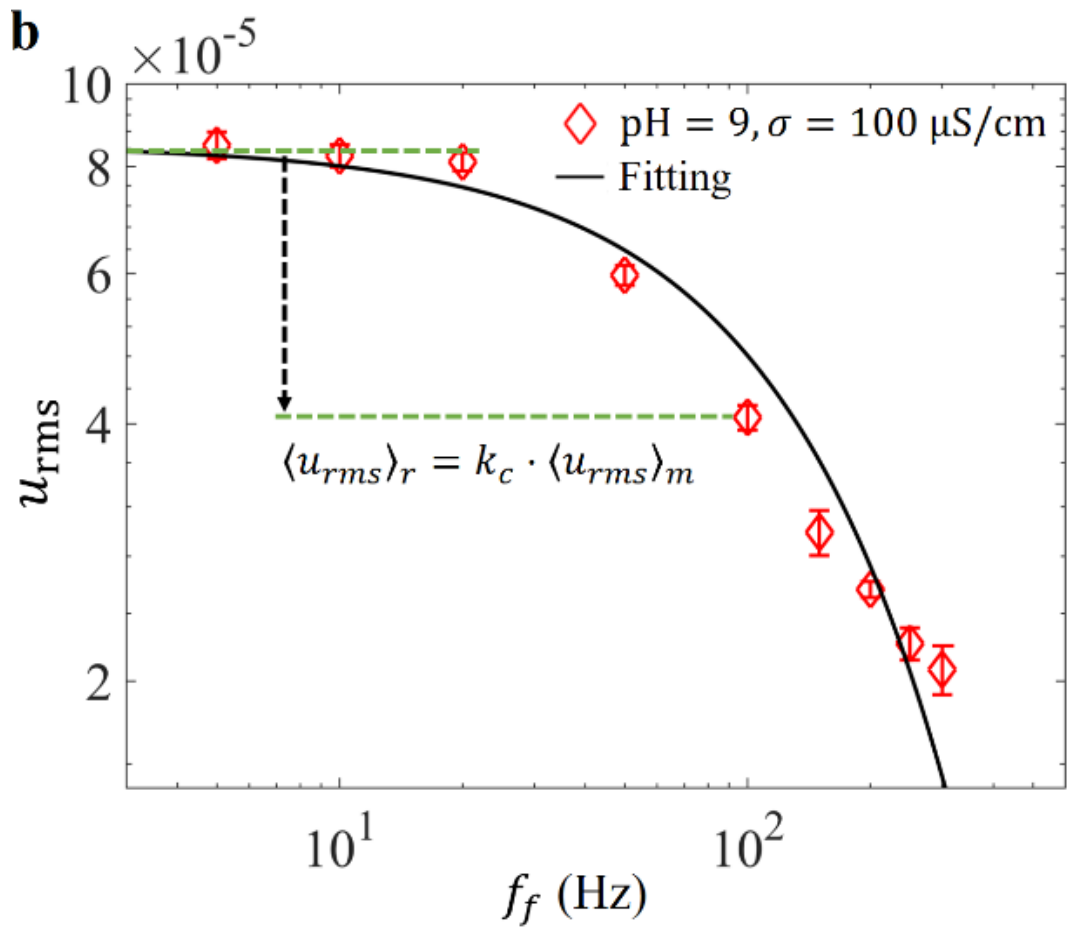

**Fig. S3 | $u_{rms}$ vs $f_f$** which is the frequency of applied AC electric field.

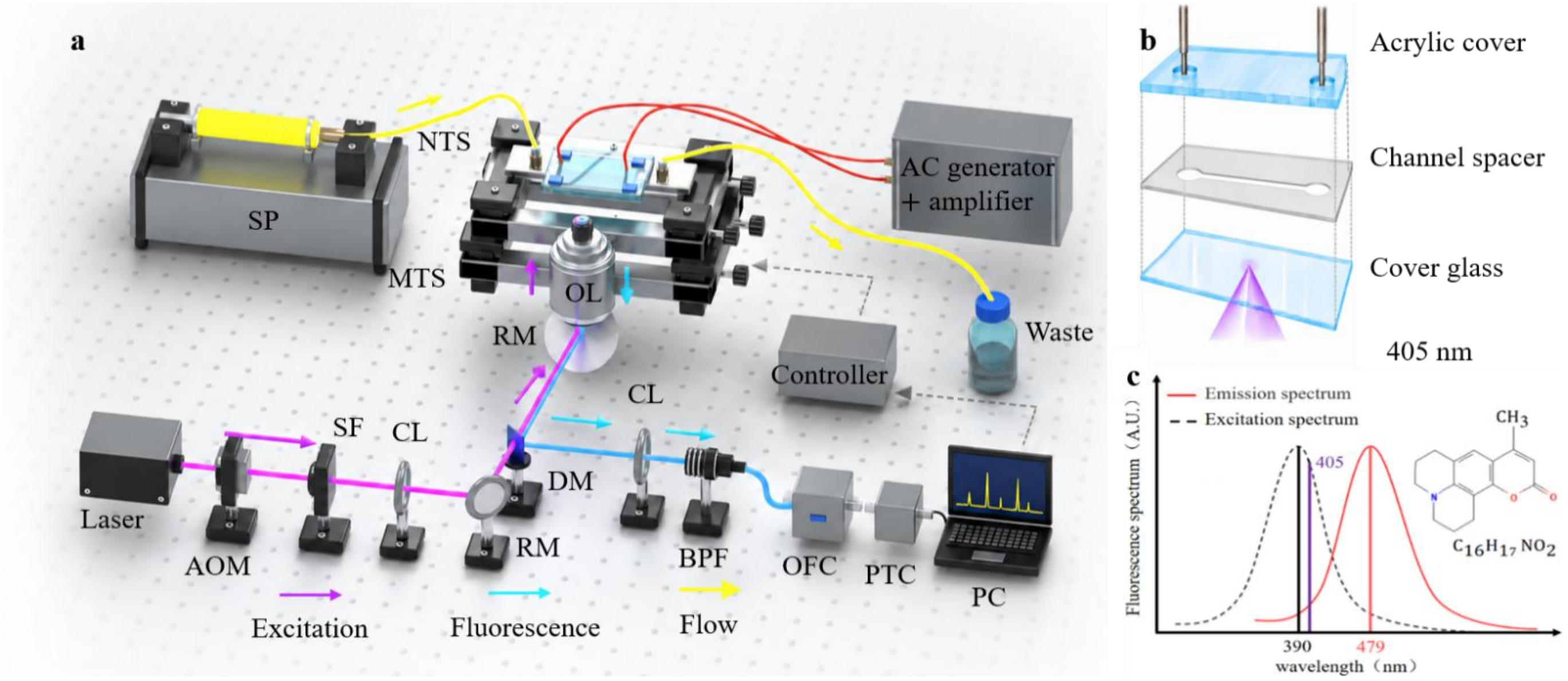

**Fig. S4 | Schematic diagram of experimental system. a,** Schematic diagram of experimental system, include: CW 405: 405 nm CW laser; AOM: acousto-optic modulator; SF: spatial filter; CL: convex lens; RM: reflective mirror; DM: dichroic mirror; OL: objective lens; BPF: bandpass filter; OFC: optical fiber connector; PTC: photon counter; MTS: micrometer translation stage; NTS: nano translational stage; SP: syringe pump; PC: computer. **b,** Schematic diagram of the microchip. **c,** Excitation and emission spectra of Coumarin 102 (Sigma Aldrich).

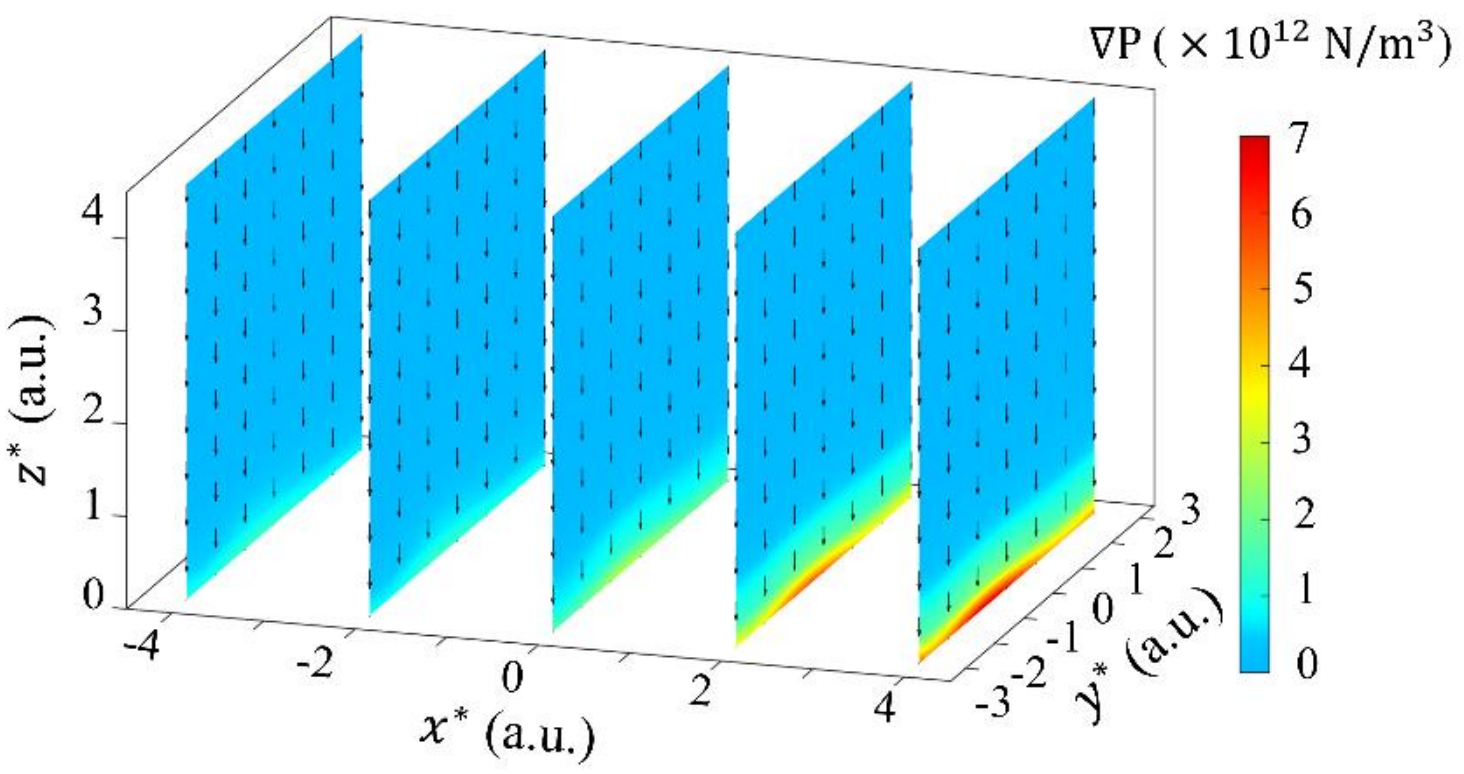

**Fig. S5 | Pressure gradient distribution.** The black arrow represents the direction of pressure gradient. The color represents the local magnitudes of pressure gradient.